\documentclass[twocolumn,secnumarabic,amssymb, nobibnotes, aps, prd]{revtex4}
\usepackage{graphicx}

\begin{document}
\title{Flux-qubit and the law of angular momentum conservation.}
\author{A.V.  Nikulov}
\affiliation{Institute of Microelectronics Technology and High Purity Materials, Russian Academy of Sciences, 142432 Chernogolovka, Moscow District, RUSSIA.} 
\begin{abstract} The confidence of many authors in the possibility to use superconducting loop interrupted by Josephson junctions as a basis for quantum bit, flux qubit, presumes the assumption on superposition of two macroscopically distinct quantum states with macroscopically different angular momentum. The contradiction of this assumption with macroscopic realism and the conservation law must call the numerous publications about flux qubit in question. These publications uncover misunderstanding by many modern physicists of the essence of the superposition principle. The Einstein - Podolsky - Rosen (EPR) correlation or entanglement, introduced in 1935 by opponents of the Copenhagen interpretation in order to reveal the contradiction of this principle with realism, has provided a basis of the idea of quantum computation. The problem of the EPR correlation has emerged thanks to philosophical controversy between the creators of the quantum theory about the subject of its description. Therefore it is impossible to solve correctly the problem of quantum computer creation without the insight into the essence of this philosophical controversy. The total neglect of the philosophical problems of quantum foundation results to concrete mistakes, the example of which are the publications about flux qubit. In order to prevent such mistakes in the future the philosophical questions about the essence of superposition and entanglement and about the subject of quantum description are considered.  
 \end{abstract}

\maketitle

\narrowtext

\section*{Introduction}

Most physicists believe in quantum mechanics because of the progress in physics and engineering of the twenty century connected with it. But John Bell in his famous Introductory remarks at Naples-Amalfi meeting, May 7, 1984 "Speakable and unspeakable in quantum mechanics" \cite{Bell1984} noted wittily that {\it "This progress is made in spite of the fundamental obscurity in quantum mechanics. Our theorists stride through that obscurity unimpeded... sleepwalking? The progress so made is immensely impressive. If it is made by sleepwalkers, is it wise to shout 'wake up'? I am not sure that it is. So I speak now in a very low voice."} Indeed, it may seem useless to pay attention to the fundamental obscurity in quantum mechanics because of the past breakthroughs. However, some modern publications force to speak loudly. 

The base of the new line of investigation of the last ten years, quantum information, quantum computation, quantum cryptography, and quantum teleportation \cite{QuIn2000} is inseparably linked with philosophical controversy \cite{QuCh2005} between the creators of the quantum theory about the fundamental obscurity in quantum mechanics. But this connection of the quantum computation problem, for example, with the unsolved problems of quantum foundation is neglected \cite{QI2008} in the most publications because of the dislike for philosophy of most modern physicists. This neglect has resulted to some delusion concerning the possibility to create quantum register \cite{QI2004} and quantum bit. For example, the authors of numerous publications about flux qubit pass over the obvious contradiction of their assumption on superposition of two macroscopically distinct quantum states of superconducting loop with macroscopic realism and even with the fundamental law of angular momentum conservation. This funny mistake will be considered in details in the first section of this paper. The intricate interrelation between quantum mechanics and the conservation laws will be considered in the second section. The numerous publications about flux qubit, unmasked as funny mistake in the first section, reveal superficial understanding by many modern physicists of the paradoxical nature of the superposition principle. The contradiction of this principle with realism and the fundamental difference of quantum mechanics from all other theories of physics are accentuated in the third section. 

\section{Flux qubit.}
The quantum computation is one of the most intriguing ideas of the last years and a practical realisation of a real quantum computer is one of the most grandiose problems. Superconductivity, one of the macroscopic quantum phenomena, is attractive for a realisation of the idea of quantum computer because of a few chance that technology will be able to work on the atomic level in the near future. Authors of numerous publications propose to made quantum bits, basic building blocks for quantum computation, on the base of superconductor structures. Quantum bit is a quantum system with two permitted states, superposition of which is possible. The superposition is the essence of the fundamental difference of quantum bit from classical one, which provides the advantage of a quantum computer over a classical one. Therefore a possibility of state superposition is the determinant of the possibility to make quantum computer on the base of superconductor structures. Many authors \cite{FlQu1, FlQu2,FlQu3,Clark08,FlQu4,FlQu5,FlQu6,FlQu7,FlQu8,FlQu9,FlQu10,FlQu11,FlQu12,FlQu13,FlQu14,FlQu15,FlQu16,FlQu17,FlQu18,FlQu19,FlQu20,Mooij03,Leggett02,Makhlin01} are sure that superconducting loop interrupted by one or three Josephson junctions can be used as quantum bit, qubit. The two permitted states superposition of which is assumed have different value of magnetic flux. Apparently therefore such loop is called flux qubit. 

\subsection{Permitted states of superconducting loop. }
The magnetic flux differs on the value $\Delta \Phi_{Ip}  = LI_{p}$ because of the persistent current $I_{p}$ circulating clockwise in the one permitted state and anti-clockwise in the other one \cite{Clark08}. Therefore the flux qubit was called also "persistent current qubit" \cite{Mooij99}. The persistent current must flow along a loop interrupted by three Josephson junctions since a stable state with zero current $I_{p} = 0$ is forbidden when magnetic flux $\Phi $ inside the loop is not divisible $\Phi \neq n\Phi _{0}$ by the flux quantum $\Phi_{0}  = \pi \hbar /e \approx  2.07 \times  10^{-15} \ T m^{2}$. The persistent current $I_{p} \neq 0$ is observed because of the demand of the current conservation (1) and a relation (2) for the phase $\phi$ of the wave function $\Psi =|\Psi |\exp{i\phi }$ describing the superconducting state in the loop. For example, in case of a loop interrupted by three Josephson junctions with critical currents $I_{c1} \approx  I_{c2} \approx  I_{c3}$ 
$$I_{p}  = I_{c1}\sin (\Delta \phi_{1}) = I_{c2}\sin (\Delta \phi_{2}) = I_{c3}\sin (\Delta \phi_{3}) \eqno{(1)} $$ 
and  
$$\Delta \phi_{1} + \Delta \phi_{2} +  \Delta \phi_{3} + 2\pi \Phi/\Phi_{0}= 2\pi n  \eqno{(2)} $$ 
The relation for the persistent current (1) results from the current-phase relationship $I_{p} = I_{c}\sin (\Delta \phi )$ between the super-current $I_{p}$ through the Josephson junction and the phase difference $\Delta \phi$ between the junction boundaries \cite{Barone} discovered in 1962 by Brian Josephson. The relation (2) is deduced from the requirement that the complex pair wave function closed in the loop must be single-valued at any point $\Psi =|\Psi |\exp{i\phi }= |\Psi |\exp{i(\phi + 2\pi n) }$. Because of this requirement the phase $\phi $ must change by integral multiples of $2\pi$  
$$\oint_{l} dl  \bigtriangledown \phi = n2\pi  \eqno{(3)}$$ 
following a complete turn along the path of integration $l$. The integral in (3) along the loop $l$, excepting Josephson junctions $\Delta \phi_{1} + \Delta \phi_{2} +  \Delta \phi_{3} $, should equal $\approx 2e\Phi /\hbar = 2\pi \Phi/\Phi_{0}$ because of the relation $\bigtriangledown \phi = p/\hbar = (mv + 2eA)/\hbar$ and the approximation $v \approx 0$, suitable for flux qubit, where $v = I_{p}/s2en_{s}$ is the velocity of superconducting pairs between the Josephson junctions.  

\subsection{Superposition is assumed between states with macroscopically different angular momentum}
Superposition is assumed between two states $n = n'$ and $n = n'+1$ having equal minimal energy at $\Phi = (n'+0.5)\Phi_{0}$ when $\Delta \phi_{1} + \Delta \phi_{2} +  \Delta \phi_{3} = 3\Delta \phi = 2\pi (n - \Phi/\Phi_{0}) = 2\pi (n - n'- 0.5) = \mp \pi $ \cite{Clark08}. Considered the limit $\beta_{L} = 2LI_{c}/\Phi_{0} \ll  1$, i.e. $\Delta \Phi_{Ip}  = LI_{p} \leq LI_{c} \ll \Phi_{0}/2$, in which the total magnetic flux $ \Phi = \Phi_{e} + \Delta \Phi_{Ip} $ equals approximately the externally applied magnetic flux $ \Phi \approx  \Phi_{e} = BS$ \cite{Clark08}.  The magnetic dependencies of the energy $E(\delta \Phi_{e}) = E(\Phi_{e}-(n'+0.5)\Phi_{0}) \propto I_{p}^{2}$ of the states $n = n'$ and $n = n'+1$ with the persistent current $I_{p,n'} = I_{c} \sin(-\pi/3 + 2\pi \delta \Phi_{e}/3\Phi_{0})$ and $I_{p,n'+1} = I_{c} \sin(\pi/3 + 2\pi \delta \Phi_{e}/3\Phi_{0})$ should cross at $\Phi = (n'+0.5)\Phi_{0}$ \cite{Clark08}. It is assumed that the energy levels can be split because of superposition of these states \cite{Clark08,Mooij00}. The experimental results \cite{ Mooij03,Mooij00} demonstrating the energy-level splitting are interpreted \cite{Clark08} as one of experimental evidences of the $n = n'$, $n = n'+1$ states superposition. There are also other observations, for example the Rabi oscillations and Ramsey interference \cite{Mooij03,Tanaka06,Semba06,Tanaka07}, which are interpreted as experimental evidence of this superposition.

Most authors are sure \cite{Clark08} that these numerous experimental evidences can not permit of a shadow of doubt about the superposition observation of the macroscopically distinct quantum states of flux qubit. Nevertheless such interpretation of the experimental results \cite{ Mooij03,Mooij00,Tanaka06,Semba06,Tanaka07} is extremely doubtful because of  some reasons.  The most obvious one is connected with macroscopic difference of the magnetic moment $M_{m} = SI_{p}$ and the angular momentum $M_{p} = (2m_{e}/e)M_{m}$ of Cooper pairs of the $n = n'$ and $n = n'+1$ states. At the values of the persistent current $I_{p} \approx  5 \ 10^{-7} \ A$ and the area $S \approx  10^{-12} \ m^{2}$ of a typical superconducting loop considered as flux qubits \cite{Mooij03} the magnetic moment equals approximately $M_{m,n'} \approx  -0.5 \ 10^{5} \ \mu _{B}$ in the $n = n'$ state and $M_{m,n'+1} \approx 0.5 \ 10^{5} \ \mu _{B}$ in the $n = n'+1$ state. Accordingly the angular momentum equals $M_{p,n'} \approx  -0.5 \ 10^{5} \ \hbar $ and $M_{p,n'+1} \approx  0.5 \ 10^{5} \ \hbar $. Where $\mu _{B}$ is the Bohr magneton and $\hbar $ is the reduced Planck constant. The spectrum of the macroscopic angular momentum and other macroscopic parameters can be discrete \cite{PRB2001} because of the same quantum number $n$, see (3), of all $N_{s} $ Copper pairs in the loop. The number of pairs $N_{s} = Vn_{s}$ is very great \cite{FPP2008} in the macroscopic volume $V \approx  ls$ of the loop, at its typical length $l = 5 \ \mu m$ \cite{Mooij03} and the section area $s \approx  0.01 \ \mu m^{2}$. Both the magnetic moment $M_{m} = SI_{p}$ and the angular momentum $M_{p} = (2m_{e}/e)SI_{p}$ have opposite directions in the $n = n'$ and $n = n'+1$ states, Fig.1, because of the persistent current circulating in the loop, clockwise in the $n = n'$ state (for example) and anticlockwise in the $n = n'+1$ state \cite{Clark08}.   

The authors of the publications about flux qubit \cite{Mooij03,Leggett02} represent superposition of the states $n = n'$, $n = n'+1$ with the relation \cite{Clark08}
$$|\Psi > = \alpha |\uparrow > + \beta |\downarrow  > \eqno{(4)} $$
by analogy with states superposition of z-projection of spin 1/2. But the angular momentum difference $M_{p,n'+1} - M_{p,n'} \approx   10^{5} \ \hbar $ between the flux qubit states $n = n'$, i.e. $ |\uparrow >$ and  $n = n'+1$, i.e. $|\downarrow  >$ is macroscopic but no microscopic $\hbar $ as in the case of the spin 1/2. The universally recognised quantum mechanics precludes up to now a possibility of superposition of states with macroscopically different angular momentum because of the contradiction of such assumption with the fundamental law of angular momentum conservation. According to the corroboration principle, introduced by Bohr as far back as 1920 \cite{Bohr1920}, and the modern quantum formalism \cite{LandauL} classical physics and quantum physics should give the same prediction of measurements results for central-symmetrical system with a macroscopic angular momentum $ \approx   10^{5} \ \hbar $. Furthermore, the quantum formalism \cite{LandauL} assumes superposition of states with different projection of angular momentum only for central-symmetrical systems, such as atom or electron. The loop, in contrast to atom or electron, is no central-symmetrical system. The persistent current circulating in the plane loop $S$ induces $M_{m} = S \times I_{p}$ and $M_{p} = (2m_{e}/e)M_{m}$ in the single direction, perpendicular to the loop plane $S$, Fig.1. Therefore the magnetic moment and angular momentum should be considered as one-dimensional in this case. Consequently, the authors of the numerous publications about flux qubit assume superposition of states of the no-central-symmetrical system with macroscopically different value of angular momentum. This assumption, contradicting not only to the conservation law but also to the universally recognised principles of quantum mechanics, can only be described as funny mistake.

\subsection{The observations of angular momentum change can not be interpreted as causeless.}
Numerous measurements \cite{Mooij03,Mooij00,1Shot02,1Shot04} of the additional magnetic flux $\Delta \Phi_{Ip}  = LI_{p}$ induced by the persistent current $I_{p}$ have shown that the probability of the flux qubit states changes between 0 and 1 in a narrow interval of magnetic field $\Delta B_{e} = \Delta \Phi_{e}/S $ near $\Phi = (n'+0.5)\Phi_{0}$, down to $\Delta B_{e} \approx 0.003 \Delta \Phi_{0}/S \approx 2 \ 10^{-7} \ T$, Fig.1, observed at $T = 0.025 \ K$ \cite{1Shot04}. These measurements means that the angular momentum of superconducting pairs changes on a macroscopic value $M_{p,n'+1} - M_{p,n'} = (2m_{e}/e)S(I_{p,n'+1} - I_{p,n'}) \approx  10^{5} \ \hbar $ at very small alteration of the magnetic field $\Delta B_{e} \approx 2 \ 10^{-7} \ T$. The interpretation of the energy-level splitting, Rabi oscillations, Ramsey interference \cite{Mooij03,Mooij00,Tanaka06,Semba06,Tanaka07} and other experimental results as experimental evidence of superposition of the $n = n'$ and $n = n'+1$ states presumes that this change can be causeless, i.e. can be observed because of macroscopic quantum tunneling \cite{Averin2008,DWave2008} without any interaction with environment. This presupposition \cite{Mooij03,Mooij00,Tanaka06,Semba06,Tanaka07} contradicts manifestly to the fundamental law of angular momentum conservation.  

\begin{figure}[]
\includegraphics{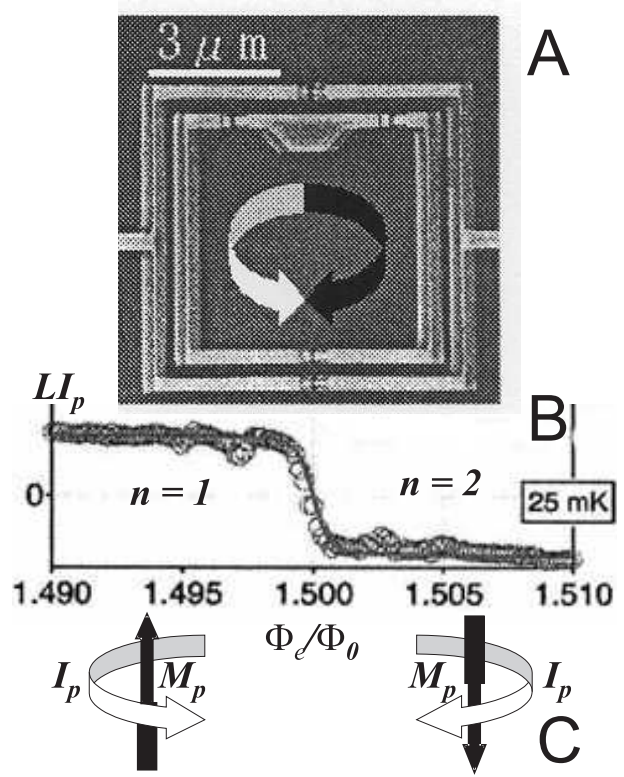}
\caption{\label{fig:epsart} A) Scanning electron microscope image of the structure used in the work \cite{1Shot04} for single-shot readout of flux qubit states is shown. The inside loop interrupted by three Josephson junctions is considered as flux qubit. The outside loop interrupted by two Josephson junctions is dc-SQUID used for measurement of the additional magnetic flux $\Delta \Phi_{Ip}  = LI_{p}$ induced by the persistent current $I_{p}$ of the flux qubit. Arrows indicate the directions of the persistent current in the flux qubit. B) The $LI_{p}$ value measured by single-shot readout and taking an average at each flux $\Phi_{e}$ value in function of external magnetic field $B_{e} = \Phi_{e}/S$. C) The change of the persistent current direction observed by means of the $LI_{p}$ sign change means the change of the angular momentum $M_{p} = (2m_{e}/e)I_{p}S$ direction.}
\end{figure}

According to the basic principles of quantum mechanics such violation of the conservation law is not possible even on the microscopic level, for spin 1/2 projection. The superposition (4) must collapse to 
$$|\Psi > = |\uparrow > \ or \  |\Psi > = |\downarrow  > \eqno{(5)} $$
after first measurement of the z-projection and all other measurements of the same projection must give the same result with probability equal to unity when any interaction with environment is absent. The superposition (4) is reinstated for the central-symmetrical system such as a particle with spin 1/2 at measurement of an other projection, different from the z-one. Such reinstatement of superposition is not possible for one-dimensional system, having single projection of angular momentum. There is important to remember that the superposition of states (4) with different values of angular momentum projection describes the uncertainty relation for measurement of perpendicular projections, along $x$, $y$ and $z$, of spin or orbital momentum of a central-symmetrical system, such as electron or atom \cite{LandauL}. It is obvious that this uncertainty relation can not be applied to one-dimensional system such as flux qubit. Therefore superposition of states with one-dimensional angular momentum can not assume even on the microscopic level, at $M_{p,n'+1} - M_{p,n'} = \hbar $. It is additional forcible argument that no experimental result obtained at measurements of flux qubit can be interpreted as experimental evidence of states superposition.

But some authors are sure not only in possibility of superposition of macroscopically distinct quantum states \cite{Clark08} and macroscopic quantum tunneling \cite{Averin2008,DWave2008} but even in a possibility to observe the superposition \cite{1Shot04}. The state with $I_{p} = 0$ is forbidden and the two permitted states $n = n'$, $n = n'+1$ with minimum energy $E_{n'} = E_{n'+1}$ have equal values but opposite directions of the persistent current $I_{p,n'} = I_{c} \sin(\pi/3 )$ and $I_{p,n'+1} = -I_{c} \sin(\pi/3 )$ at $\Phi = (n'+0.5)\Phi_{0}$ because of the demands of the current conservation (1) and quantization (2). Therefore a single-shot measurement of the persistent current must give values $I_{p,n'} = I_{c} \sin(\pi/3 ) \neq 0$ or $I_{p,n'+1} = -I_{c} \sin(\pi/3 ) \neq 0$ whereas this value taking an average must be equal zero $\overline{I_{p}} = P_{ n'+1}I_{p,n'+1} + P_{ n'} I_{p,n'} $ since the probability $ P_{ n'} \approx e^{-E_{n'}}/(e^{-E_{n'}}+ e^{-E_{n'+1}})$ equals 0.5 at $\Phi = (n'+0.5)\Phi_{0}$, when $E_{n'} = E_{n'+1}$. The probability $ P_{ n'}$ changes with magnetic field $\Delta B_{e} = \Delta \Phi_{e}/S$ because of the energy $E_{n'}$, $E_{n'+1}$ change \cite{Clark08}. The single-shot measurement should give the two parallel lines of the dependencies $ LI_{p,n'}(\delta \Phi_{e})$, $LI_{p,n'+1}(\delta \Phi_{e})$ in a narrow interval $\delta \Phi_{e} = \Phi_{e}-(n'+0.5)\Phi_{0} \ll  \Phi_{0}$ where both $ P_{ n'}$ and $ P_{ n'+1}$ have a noticeable value. Such parallel lines, corresponding to quantum mechanics prediction, are observed in \cite{1Shot02}, Fig.4,  and \cite{1Shot04} Fig.4b. But the authors \cite{1Shot04} interpret such dependencies as {\it the classical behaviour of a two-level system}. They observed in different samples a $\chi $-shaped crossing of the dependencies $LI_{p,n'}(\delta \Phi_{e})$, $LI_{p,n'+1}(\delta \Phi_{e})$ in different samples. The authors \cite{1Shot04} interpret such behaviour as observation of {\it the coherent superposition of macroscopic supercurrent states}. Obviously, such interpretation have nothing in common with the universally recognised quantum mechanics. Each physicists must understand that it is impossible to observe a superposition even of microscopic states. In spite of the misunderstanding of quantum mechanics by the authors \cite{1Shot04} their experimental results are very interesting. The decrease of the interval $\delta \Phi_{e} $ of the average value $\overline{I_{p}}$ change  with temperature decrease, see Fig. 5 in \cite{1Shot04}, corroborates that the transition between the $n = n'$ and $n = n'+1$ states takes place because of an interaction with environment.

\subsection{A compensation of the angular momentum change can not be assumed. } 
Some authors understand that the conservation law must prohibit quantum tunnelling between states with different magnetic moment and angular momentum. The author of \cite{Chudnov} proposes to assume a firm coupling with a large solid matrix ($lsm$) that absorbs the change in the angular momentum. Neither angular momentum conservation problem nor such firm coupling are consider in the numerous publications on superposition of flux qubit states \cite{Makhlin01,Mooij99,Leggett2002} and macroscopic tunneling \cite{Averin2008,DWave2008}. Nevertheless my conversations with some authors of these publications have shown their hope that the change of the angular momentum of superconducting condensate ($sc$) observed at the transition between $n = n'$ and $n = n'+1$ states can be compensated with a change of an angular momentum of a large solid matrix, i.e. the loop, substrate and so forth. This hope presumes the entanglement 
$$\Psi_{qubit} = \alpha |\uparrow >_{sc}|\downarrow  >_{lsm} + \beta |\downarrow  >_{sc} |\uparrow >_{lsm}     \eqno{(6)} $$
of the states of superconducting condensate $|\uparrow >_{sc}$, $|\downarrow  >_{sc}$ and a large solid matrix $|\downarrow  >_{lsm}$, $|\uparrow >_{lsm}$ of uncertainly large mass. Thus, following to \cite{Chudnov} the authors of  \cite{Makhlin01,Mooij99,Leggett2002, Averin2008,DWave2008} and others publications on flux qubit must assume that not only superconducting condensate but also an uncertain system with macroscopic mass can be in superposition of states.  It is doubtful that the fantasy about superposition (6) of the loop, substrate and so forth states $|\downarrow  >_{lsm}$, $|\uparrow >_{lsm}$ could be taken seriously even by the authors of publications about flux qubit.

Therefore the superconducting loop interrupted by Josephson junctions can not be considered as flux qubit. The energy-level splitting, Rabi oscillations, Ramsey interference and other experimental results \cite{Mooij03,Mooij00,Tanaka06,Semba06,Tanaka07} must be explained without using of the superposition principle. We must explain the transition between the states $n = n'$ and $n = n'+1$ of superconducting loop observed at measurement of the $\Delta \Phi_{Ip}  = LI_{p}$ values \cite{Mooij03,Mooij00,1Shot02,1Shot04} as a consequence of an interaction with environment or an external influence, in order that we can avoid the contradiction with the conservation law and the universally recognised principles of quantum mechanics.

\section{Intricate interrelation between quantum mechanics and the conservation laws}
In view of the contradiction the flux qubit assumption with the law of angular momentum conservation it is urgent to consider the interrelation between quantum mechanics and the conservation laws which is enough intricate. It can not be precluded that the authors of the numerous publications about flux qubit may claim that the difficulty with conservation law is peculiar no only flux qubit assumption but it is just one of several paradoxes associated with the concept of "measurement" in quantum mechanics. The creators of the quantum mechanics tried to avoid any contradiction with the conservation laws and some aspects of quantum formalism are closely associated with these laws \cite{LandauL}. Therefore many modern physicists believe that there is no problem with conservation laws in quantum mechanics. In contrast to this belief N. Bohr emphasized the contradiction between a realistic description of some quantum  phenomena and the conservation laws: {\it "Any attempt of an exact space-time description of quantum phenomena implies the refusal of strict application of the conservation law. And vise versa, a strict application of the conservation laws in quantum phenomena implies an essential restriction of exactness of the space-time description"} \cite{Bohr1936}.  

In order to remove the obvious problems with conservation laws inside the quantum-mechanical formalism its creators, Bohr, Heisenberg and others have swept these problems out  this formalism, in the concept of measurement, which is most vague in quantum mechanics. The obvious vagueness of this concept was accentuated for example by John Bell: {\it "The concept of 'measurement' becomes so fuzzy on reflection that it is quite surprising to have it appearing in physical theory at the most fundamental level"} \cite{Bell1987}. He has entitled one of his last papers "Against Measurement"  \cite{Bell1990}: {\it "Here are some words which . . . have no place in a formulation with any pretension to physical precision: system, apparatus, environment, microscopic, macroscopic, reversible, irreversible, observable, information, measurement. On this list of bad words the worst of all is "measurement""}. 

The concept of 'measurement' is reduced virtually to the words about the collapse of superposition (wave function) or {\it a "quantum jump" from the 'possible' to the 'actual' which must take place during the act of observation} \cite{Heisenberg1958}. The collapse of superposition postulated by von Neumann in 1932 \cite{Neumann1932} is formally outside of quantum-mechanical formalism. But without this postulate the superposition principle can not be applied for the description of quantum phenomena. We can not observe different results of single observation simultaneously. Einstein, Podolsky Rosen have prove unambiguously in their famous paper \cite{EPR1935} that a description of physical reality with help of superposition and its collapse at observation can be considered complete only if non-local interaction is possible in this reality. The contradiction of the superposition principle with realism will be considered in the next section. This section is devoted to the problem with the conservation law in quantum phenomena. It will be shown that the universally recognised quantum mechanics admits some causeless change of parameters at observation in spite of the conservation law, but restricts the possibility of such change within the limits of the uncertainty relation and the Planck's constant $\hbar $.

\subsection{The challenge to the conservation law in the Stern-Gerlach effect.} 
The Stern-Gerlach effect was the utmost importance in the history of quantum mechanics and the controversy about its foundation. The paradoxical nature of this effect was realised \cite{Einstein1922} just after its discovery in 1922. Bohr wrote about that time \cite{Bohr1949}: {\it "In the following years, during which the atomic problems attracted the attention of rapidly increasing circles of physicists, the apparent contradictions inherent in quantum theory were felt ever more acutely. Illustrative of this situation is the discussion raised by the discovery of the Stern-Gerlach effect in 1922. On the one hand, this effect gave striking support to the idea of stationary states and in particular to the quantum theory of the Zeeman effect developed by Sommerfeld; on the other hand, as exposed so clearly by Einstein and Ehrenfest \cite{Einstein1922}, it presented with unsurmountable difficulties any attempt at forming a picture of the behaviour of atoms in a magnetic field"}. Indeed, it seems impossible to describe realistically the experimental results, obtained first by Stern and Gerlach \cite{SternGerlach}, according to which magnetic moment has an identical value of projection on any direction. Therefore the Stern-Gerlach effect is considered as the main example in the publications  \cite{QuCh2005,Neumann1932,Bohm1951,Bell1966,Bell1964,Mermin1993} devoted to the controversy about a possibility of a realistic description of quantum phenomena. The Stern-Gerlach effect challenges the law of angular momentum conservation, as well as the flux qubit assumption. Therefore it is useful to consider just this effect in this paper. 

\begin{figure}
\includegraphics{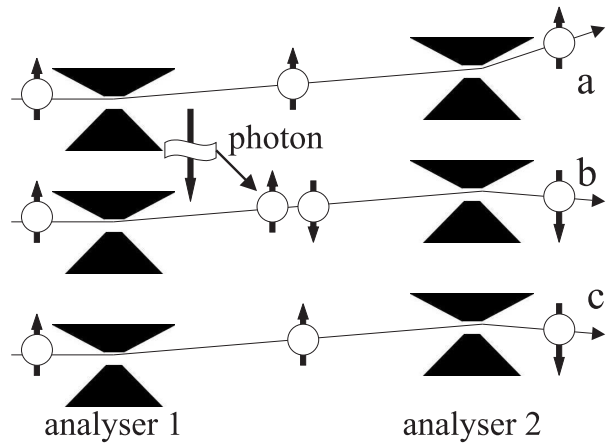}
\caption{\label{fig:epsart} a) Two consecutive measurements of a spin projection must give the identical results when the axes of the first (analyser 1) and second (analyser 2) Stern-Gerlach analysers are exactly parallel and any interaction with environment is absent. b) The results of these measurements can be contrary if a photon has compensated the angular momentum change observed in this case. c) According to the orthodox quantum mechanics the contrary results can be observed without any compensation of the angular momentum change when the axes of the analysers are no exactly parallel. }
\end{figure}

The Stern-Gerlach experiment has revealed that all particles with spin 1/2 deviate from their initial trajectory because of a gradient of magnetic field only on two angles corresponding to the two values $\mu _{B}$ and $-\mu _{B}$ of magnetic moment projection. The axis of the Stern-Gerlach analyser, along which the magnetic field gradient is directed, can have any orientation. Consequently, measurements of angular momentum projection of particles with spin 1/2 in any directions give only two values $\hbar /2$ and $-\hbar /2$. The orthodox quantum mechanics describes the paradoxical results of the Stern-Gerlach experiment with help of the superposition (4) of the two possible results of measurements and its collapse to one of the eigenstates (5) at observation. According to the quantum formalism \cite{LandauL} the eigenstate of the z-projection, for example, 
$$\Psi_{spin-1/2} = |\uparrow >_{z}  \eqno{(7)} $$
is superposition 
$$\Psi_{spin-1/2} = cos(\theta) |\uparrow >_{\theta } + sin(\theta) |\downarrow  >_{\theta }  \eqno{(8)} $$ 
for a $\theta $ projection along the direction turned from z-one on a angle $\theta $. Therefore after the collapse of the superposition (4) to the eigenstate (7) at z- projection measurement with help of the first Stern-Gerlach analyser all posterior measurement of z-projection must give the same result with the probability $1^{2} = 1$, Fig.2a. If the opposite result will be observed we should conclude that an interaction with environment takes place, Fig.2b. For example a photon has compensated for the angular momentum change which we observed, so that total angular momentum has been conserved. But result of measurement can change without photon if we will turn the axis of the second Stern-Gerlach analyser, i.e. the direction of its magnetic field gradient, on a angle $\theta $. The eigenstate of the z-projection (7) is superposition (8) for the $\theta $ projection along the new direction of the analyser axis. Therefore after the result spin up $|\uparrow >$ at measurement on the first Stern-Gerlach analyser we can obtained with the probability $ sin^{2}(\theta)$ the result spin down $|\downarrow  >$ at measurement on the second Stern-Gerlach analyser, Fig.2c. 

Thus, the orthodox quantum mechanics predicts the observation of the causless change of the angular momentum projection without photon and any other interaction with environment in defiance of the conservation law. There is important to note that the contradiction with the fundamental law is predicted for the results of observations. The relative number $\propto  sin^{2}(\theta)$ of particles  for which this contradiction should be observed decreases with the turn angle $ sin^{2}(\theta) \approx  \theta ^{2}$. But the violation of the conservation law is the violation even if it is observed for only particle which could be at almost parallel orientation $\theta \approx 0$ of the axes of the first and second analysers, Fig.2c. The adherents of the Copenhagen interpretation may state that this contradiction with the conservation law can avoid with help of the quantum postulate proposed by Bohr in 1928 \cite{Bohr1928} according to which {\it any observation of atomic phenomena should include an interaction they with equipment used for the observation which can not be neglected}. But it is mere words in which the adherents can believe. No one of they did not explain how the Stern-Gerlach analyser, for example, can adsorb the angular momentum change which should be observed in the experiment shown on the Figure 2c.

\subsection{Spin 1/2 and flux qubit.}
The challenge to the conservation law observed in the Stern-Gerlach effect can not apply to the case of flux qubit. Both this challenge and using of superposition principle for the description of the Stern-Gerlach effect are closely associated with the paradoxicality of the observations of the same value of the angular momentum projection measured in any direction. This paradoxicality can not be observed logically in the case of flux qubit because of its one-dimensionality. Therefore only unfounded assumption on superposition but no observation can challenge to the law of angular momentum conservation in this case. It is enough to decline this assumption in order to remove this challenge to the fundamental law. We must do this in the case of flux qubit in order to do not contradict to the universally recognised principles of quantum mechanics. According to these principles any challenge to the conservation laws at observation must be in the limit of the uncertainty relations and the causless change of the angular momentum can not exceed considerably the $\hbar $ value. Just the uncertainty relations allow to do not contradict to the conservation law at consideration of some quantum phenomena, according to Bohr: {\it "Any strong application of the laws of momentum and energy conservation to atomic processes presupposes the refusal of quite certain localisation of particle in space and time. This circumstance is represented quantitatively with the Heisenberg's uncertainty relation"} \cite{Bohr1958}.

\section{Why could superposition of flux qubit states be assumed?}
The clockwise and anticlockwise direction of the persistent current in the $n = n'$ and $n = n'+1$ states is assumed \cite{Clark08} and observed \cite{1Shot02,1Shot04}. Therefore numerous authors of publications about flux qubit must realise that these states of superconducting loop have macroscopically different angular momentum. Then why could they believe in the superposition of these states in spite of the obvious contradiction with the fundamental law of angular momentum conservation? My conversations with some authors of publications about flux qubit allow to suppose that this funny mistake could become possible because of their disregard of philosophical problems of quantum mechanics foundation.  

According to L.D. Landau \cite{LandauL} and other adherents of the Copenhagen interpretation the superposition of states is the cardinal positive principle of quantum mechanics. Most physicists have already got accustomed to this principle in the course of eighty years history of quantum mechanics and they are well-informed about superconductivity as one of the macroscopic quantum phenomena. It may be therefore the possibility of superposition of states of some superconductor nanostructures seems self-evident for many physicists who do not realise enough profoundly the essence of the Bell's statement on {\it the fundamental obscurity in quantum mechanics}. The funny mistake with flux qubit and some other mistakes connected with the idea of quantum computations force to speak loudly about this fundamental obscurity. In order to understand the essence of this obscurity it is useful to consider the controversy about the subject of the quantum mechanics description. This controversy was begun by the creators of quantum theory and is still in progress up to now. 

\subsection{What can the superposition principle describe? } 
The quantum mechanics, the cardinal positive principle of which is the superposition of states, was created on the base of the proposal of Werner Heisenberg made in 1925 \cite{Heisenberg1925} to consider only observable parameters and to do not consider hidden variables. In fact, Heisenberg has proposed that quantum theory should describe only phenomena, which we can observed, but no reality, which can exist irrespective of any act of observation. Such theory can be considered as complete description only according to the positivism point of view that it makes no sense to say about a reality without an observer. Heisenberg defended just this positivism point of view. He wrote in 1958 \cite{Heisenberg1958}: {\it "In classical physics science started from the belief - or should one say from the illusion? - that we could describe the world or at least parts of the world without any reference to ourselves"}. 

Einstein forewarned as far back as 1928 that the positivism of Heisenberg and Bohr may result to the sleepwalking, about which Bell said in 56 years later. He wrote to Schrodinger  \cite{Lett1928}: {\it "The soothing philosophy-or religion?-of Heisenberg-Bohr is so cleverly concocted that it offers the believers a soft resting pillow from which they are not easily chased away"}. Einstein, as well as Bell, {\it fully recognised the very important progress which the statistical quantum theory has brought to theoretical physics} \cite{Einstein1949}. But he was {\it "firmly convinced that the essentially statistical character of contemporary quantum theory is solely to be ascribed to the fact that this [theory] operates with an incomplete description of physical systems"} \cite{Einstein1949}. He adduced in 1949 \cite{Einstein1949} {\it reasons which keep he from falling in line with the opinion of almost all contemporary theoretical physicists}: {\it "What does not satisfy me in that theory, from the standpoint of principle, is its attitude towards that which appears to me to be the programmatic aim of all physics: the complete description of any (individual) real situation (as it supposedly exists irrespective of any act of observation or substantiation)"}. Arguing against the positivism point of view of Heisenberg, Bohr and other adherents of the Copenhagen interpretation Einstein persisted that {\it "it must seem a mistake to permit theoretical description to be directly dependent upon acts of empirical assertions, as it seems to me to be intended [for example] in Bohr's principle of complementarity"} \cite{Einstein1949}. The superposition principle, in accordance with the positivism point of view of the Copenhagen interpretation, can describe only results of observation, but no a real process or situation.

\subsection{What do we observe?} 
The words by Einstein about {\it the soothing philosophy of Heisenberg-Bohr} as {\it a soft resting pillow } turned out prophetic. Already some generations of physicists studied quantum mechanics in its Copenhagen interpretation. But most of their did not understand that the subject of its description differs in essence from the one of other theories of physics.  Richard Feynman wrote in 1967: {\it "I think it is safe to say that no one understands quantum mechanics"} \cite{Feynman1967}. Feynman was right in the sense that we do not know what could manifest the quantum phenomena so perfectly described by quantum mechanics. The quantum mechanics describes results of our observation. But it can not answer unambiguously on the question: "What do we observe?" Therefore quantum mechanics, in contrast to other theories of physics, has many interpretations \cite{WikiInter}, which propose different answers on this question or refuse to answer on it. The Copenhagen interpretation is based just on such refusal. There is important to understand that it is impossible to conclude about possibility of a real quantum computer in the limits of the Copenhagen interpretation because of this refusal. A real equipment can not be made on base of phenomena description if these phenomena are not manifestation of an objective reality. In order to decide against or in favour of real quantum computer we must use an interpretation, which proposes an answer on the question: "What do we observe?" Unfortunately our decision should depend on our choice of interpretation. According to some interpretations, for example the many-worlds interpretation \cite{Everett} and non-local hidden variables interpretation by David Bohm \cite{Bohm1952}, real quantum computer is possible whereas according to the other one, for example the statistical interpretation \cite{Ball1970,Ball1998}, it is not possible.  

The ideas of David Deutsch and Richard Feynman are considered \cite{DiVincenzo} as the original sources of the numerous publications about quantum computer. It is important to note that the ideas of Deutsch and Feynman are different in their essence. Deutsch invented the idea of the quantum computer in the 1970s as a way to experimentally test the "Many Universes Theory" of quantum physics - the idea that when a particle changes, it changes into all possible forms, across multiple universes \cite{Father}. This theory is one of the realistic interpretations \cite{Everett} of quantum mechanics which allows to interpreted most paradoxical quantum phenomena without external observation, as manifestation of real processes. But this processes should occur across multiple universes \cite{DeutschFR}. According to Deutsch, {\it "quantum superposition is, in Many Universes terms, when an object is doing different things in different universes"} \cite{Father}. The Many Universes interpretation allows to understand why quantum computer may excel the classical one. It can do {\it "a number of computations simultaneously in different universes"} \cite{Father}. But the idea of many Universes seems mad for most physicists. Therefore most authors follow to Richard Feynman and consider the idea of quantum computing in the limits of the Copenhagen interpretation. 

But this idea can not be understood correctly in these limits. Heisenberg warned against the incorrect interpretation of quantum mechanics: {\it "A real difficulty in the understanding of the Copenhagen interpretation arises, however, when one asks the famous question: But what happens 'really' in an atomic event?"} \cite{Heisenberg1958}. He emphasized many times that {\it "there is no way of describing what happens between two consecutive observations"} and {\it "that the concept of the probability function does not allow a description of what happens between two observations"} \cite{Heisenberg1958}. But in spite of these warnings most modern authors are fully confident that the probability function (i.e. the Schroringer wave function or $\Psi $-function) can describe the real process of quantum computation although it should be just between observations.

\subsection{Superposition versus realism.} 
There is important to note that the basic principle of the idea of quantum computation was introduced in 1935 by opponents of the Copenhagen interpretation. Both Einstein, Podolsky, Rosen \cite{EPR1935} and Schrodinger where sure that this principle, entanglement or Einstein - Podolsky - Rosen correlation, can not be real because of its contradiction with locality principle. Therefore Einstein, Podolsky, Rosen stated that quantum- mechanical description of physical reality can not be considered complete \cite{EPR1935} and Schrodinger introduced \cite{Schrod35} this principle as "entanglement of our knowledge" \cite{Entangl}. The EPR paradox \cite{EPR1935} has prove unambiguously that a description of physical reality with help of superposition principle can be considered complete only if non-local interaction is possible in this reality. In the Bohm's version \cite{Bohm1951} of the EPR paradox the spin states of two particles are entangled  
$$\Psi_{EPR} = \alpha |\uparrow >_{A}|\downarrow  >_{B}   + \beta |\downarrow  >_{A}|\uparrow >_{B} \eqno{(9)} $$
because of the law of angular momentum conservation. Any measurement of spin projection must give opposite results independently of the distance between the particles $r_{A} - r_{B}$ because of this fundamental conservation law. The description of this correlation with help of superposition and its collapse  
$$\Psi_{EPR} = |\uparrow >_{A}|\downarrow  >_{B} \eqno{(10)} $$
implies that a measurement of the particle $A$ can instantly change a state of the particle $B$. The correlation between results of measurements of spin projections of the particles $A$ and $B$ predicted with orthodox quantum mechanics must reveal a real non-locality if superposition (9) is interpreted as description of a reality.

Many modern authors following to the book \cite{Chuang2000} interpret violation of the Bell's inequality \cite{Bell1964} as an experimental disproof of the EPR statement on incompleteness of quantum description of physical reality. Using Bell's terminology I must say that this interpretation is a consequence of the sleepwalking, which in the present case results to no immensely impressive progress but to mass delusion. The results of the Bell's experiments \cite{EPRexper} give evidence rather of the physical reality absence than of the EPR blunder. There is important to understand that the EPR correlation is quite incompatible with local realism in its essence. It has revealed unambiguously intrinsic non-locality of superposition principle in description \cite{EPR1935}. The Bell's experiments \cite{EPRexper} give evidence of the EPR correlation in quantum phenomena and has revealed non-locality in observation. These two kind of non-locality were distinguished in 1986  \cite{Cramer1986} as non-locality of the first kind and non-locality of the second kind. The authors \cite{Martinis2009} claim already on observation of Bell's inequality violation in a superconductor structure, i.e. on non-locality of the second kind in the macroscopic sphere. This unfounded claim on violation of macroscopic realism \cite{Leggett1985} is one more consequence of  the sleepwalking equally with flux qubit.

David Mermin quotes in the paper "Hidden variables and the two theorems of John Bell" \cite{Mermin1993} {\it a celebrated polymath} who declared that {\it most theoretical physicists} do not distinguish between that what is not measurable and that what is unreal and nonexistent. This remark reveals the main cause of the incomprehension by {\it most theoretical physicists} of the philosophical importance of the Bell's works \cite{Bell1964,Bell1966} and of the philosophical aspect of the quantum computer problem. Mermin has divided as far back as  1985 \cite{Mermin1985} the physicists concerning their attitude to violation of the Bell's inequalities into a majority who are "indifferent" and a minority who are "bothered". The first, for example the authors \cite{Chuang2000}, are sure that violation of the Bell's inequalities proves only that Einstein was not right and we can continue to sleep on the {\it soft resting pillow} offered by Heisenberg and Bohr. Whereas the second, for example the authors \cite{Cramer1986}, understand that it is experimental evidence of non-locality or absence of reality, i.e. violation of local realism in the observation. 

The minority bother about contradiction between the principle of superposition and realism. This contradiction is accentuated with the title "Is the moon there when nobody looks? Reality and the quantum theory" of the paper \cite{Mermin1985}. The question about moon, first raised by Einstein in 1950 \cite{Pais1982}, should emphasise that orthodox quantum mechanics using the superposition principle does not assume a real existence of  a parameter before its measurement. Therefore the superposition principle can be acceptable only in the positivism approach according to which physical theory should describe no real processes but only results of observation. Just this refusal to consider {\it any real situation} does not satisfy Einstein in the orthodox quantum mechanics. Einstein persisted that the quantum mechanics describing only phenomena but no objective reality can not be considered complete theory. Bell shared this dissatisfaction by Einstein. He accentuated that quantum mechanics refusing to describe parameters of quantum system can describe only parameters of measuring apparatus. He connected the fundamental obscurity in quantum mechanics with the problem: {\it "how exactly is the world to be divided into speakable apparatus. . .that we can talk about. . .and unspeakable quantum system that we can not talk about?"} Bell as well as Heisenberg, Einstein and other creators of quantum theory understood that orthodox quantum theory can not describe quantum reality. Nevertheless most modern authors are fully confident that it can describe a process of quantum computation.    

The paper "Quantum mechanics versus macroscopic realism: Is the flux there when nobody looks?" \cite{Leggett1985} witnesses that Anthony Leggett, who assumed first a superposition of two macroscopically distinct quantum states of a superconducting loop \cite{Clark08}, understands the contradiction between superposition and realism. According to the title and the essence of the paper \cite{Leggett1985} the superconducting loop can be used as flux qubit if only the flux $\Delta \Phi_{Ip}  = LI_{p}$ does not exist when nobody looks, i.e. if no realistic description is possible for some quantum phenomena observed at measurements of this loop. Such impossibility of realistic description is revealed with help of so called {\it no-go theorem} or {\it no-hidden-variables theorem} \cite{Mermin1993}. The authors \cite{Leggett1985} propose a no-go theorem which, as they state, similar to the famous one by John Bell \cite{Bell1964}. But, as L. E. Ballentine notes \cite{Ball1987}, the analogy between the Bell-type inequalities and the Leggett-Garg inequalities can be misleading. A key role is played in the Bell's no-go theorem by the locality postulate \cite{Bell1964}, which can not be applicable to the single localised system, i.e. single superconducting loop, considered by Leggett and Garg \cite{Leggett1985}. The locality requirement is a decisive factor of the no-go theorems. John Bell pointed out the hidden variable interpretation of David Bohm \cite{Bohm1952} as example of a non-local theory reproducing all observation prediction given by quantum formalism. He has shown \cite{Bell1966} that von Neumann's no-go proof \cite{Neumann1932}, which does not used the locality requirement, was based on an unreasonable assumption \cite{Mermin1993}. Bell has constructed \cite{Bell1966} a hidden-variables model for a single spin 1/2 that reproduces all predictions of measurement results given by the orthodox quantum theory \cite{Mermin1993} in spite of paradoxical result of the Stern-Gerlach experiment. This Bell's disproof of the von Neumann no-go theorem means that no experimental results, even the paradoxical Stern-Gerlach one, obtained on a single two-state system can contradict to realism and give evidence of state superposition. Therefore it is a consequence of sleepwalking that many authors venture to interpret experimental results obtained on flux qubit, two-states system with one-dimensional angular momentum, as an evidence of superposition of macroscopic states.

\section*{Conclusion} 
The obvious contradiction with the fundamental law of angular momentum conservation and with the universally recognised quantum formalism must unambiguously manifest that the numerous publications about flux qubit are a consequence of a funny mistake. This mistake has become possible because of the neglect by many modern authors of the "philosophical" question on the object of quantum-mechanical description. In contrast to most modern physicists the creators of quantum theory well realised the importance of epistemological problems and devoted considerable energy to their solution. They proposed different solutions, but, what is important, both Einstein and Bohr with Heisenberg understood epistemological problems of quantum mechanics. The story of flux qubit manifests that these problems have not only philosophical but also practical importance. It may be important for a solution of the practical problem of quantum computer creation to remind the Einstein opinion about the relationship of epistemology and science: {\it "The reciprocal relationship of epistemology and science is of noteworthy kind. They are dependent upon each other. Epistemology without contact with science becomes an empty scheme. Science without epistemology is - insofar as it is thinkable at all - primitive and muddled. However, no sooner has the epistemologist, who is seeking a clear system, fought his way through to such a system, than he is inclined to interpret the thought-content of science in the sense of his system and to reject whatever does not fit into his system. The scientist, however, cannot afford to carry his striving for epistemological systematic that far. He accepts gratefully the epistemological conceptual analysis; but the external conditions, which are set for him by the facts of experience, do not permit him to let himself be too much restricted in the construction of his conceptual world by the adherence to an epistemological system. He therefore must appear to the systematic epistemologist as a type of unscrupulous opportunist: he appears as realist insofar as he seeks to describe a world independent of the acts of perception; as idealist insofar as he looks upon the concepts and theories as the free inventions of the human spirit (not logically derivable from what is empirically given); as positivist insofar as he considers his concepts and theories justified only to the extent to which they furnish a logical representation of relations among sensory experiences. He may even appear as Platonist or Pythagorean insofar as he considers the viewpoint of logical simplicity as an indispensable and effective tool of his research"} \cite{Einstein1949}.

\section*{Acknowledgments} 
This work has been supported by a grant "Possible applications of new mesoscopic quantum effects for making of element basis of quantum computer, nanoelectronics and micro-system technic" of the Fundamental Research Program of ITCS department of RAS and the  Russian Foundation of Basic Research grant 08-02-99042-r-ofi.

\end{document}